\documentclass[12pt]{iopart}
\usepackage{graphicx}

\begin{document}

\title
[Superconductivity in metallic nanofilms]
{Influence of the electron density on the thickness-dependent energy gap oscillations in 
superconducting metallic nanofilms}

\author
{P W\'ojcik$^1$, M Zegrodnik$^{1}$}

\address
{$^1$AGH University of Science and Technology, Faculty of
Physics and Applied Computer Science, 
al. A. Mickiewicza 30, Krak\'ow, Poland}
\address
{$^2$AGH University of Science and Technology, Academic Centre for Materials and Nanotechnology,
al. A. Mickiewicza 30, Krak\'ow, Poland}

\ead{pawel.wojcik@fis.agh.edu.pl}


\submitto{Journal of Physics: Condensed Matter}

\maketitle

\begin{abstract}
The thickness-dependent energy gap oscillations in the metallic nanofilms are investigated by the use 
of the self-consistent numerical solutions of the Bogoliubov-de Gennes equations. 
It is shown, that the oscillations are induced by the quasi-particle energy quantization triggered by the confinement 
of electrons in the direction perpendicular to the sample. 
We have analyzed, how the changes in the electron density of states ($n_e$) and the electron-phonon coupling constant ($g$) 
influence the amplitude of the considered oscillations. It has been found, that the increase in $n_e$ and the decrease in $g$, 
can lead to a significant reduction of the oscillations amplitude. As a result, for the values of the mentioned 
parameters corresponding to some of the realistic situations the thickness-dependent 
superconducting gap oscillations can be almost completely suppressed. 
\end{abstract}

\section{Introduction}\label{sec:intro}

The huge progress in nanotechnology which has been made in the last decade 
allows for the fabrication of high quality metallic nanostructures, e.g.,
metallic nanowires~\cite{Zgirski2005,Tian2005,Altomare2006,Jankovic2006} 
and nanofilms with thickness of few monolayers~\cite{Guo2004,Bao2005,Eom2006,Ozer2006,Ozer2006_2,Ozer2007,Zhang2010,Qin2009}.
With this respect superconducting nanofilms have attracted growing interest during recent years. 
One should also note that studies have been made regarding superconducting superlattices for both 
conventional (BCS)~\cite{Frick1992} and high-Tc materials~\cite{Spalek1996}.
When the miniaturization reaches the level for which the size of
the nanostructure becomes smaller then the phase coherence length $\xi$, the
superconducting properties of the system start to deviate significantly from
those in the bulk. The distinctive experimental manifestation of the 
size effect is the non-zero resistance below $T_C$ of the quasi-1D superconducting wires~\cite{Lau2001,Bezryadin2000,Tian2005}.
Such behavior results from the phase fluctuations which occur when the diameter of
the nanowire is reduced to few tens of nanometer, i.e., thermally activated phase slip~\cite{Langer1967,McCumber1970} close to
$T_c$ and quantum phase slip~\cite{Giordano1990} at temperature far below $T_c$.
In the nanoscale regime the superconducting properties of the system change also due to the
simple fact, that the reduction of the electron motion results 
in the quantization of its energy. In consequence, the Fermi sphere splits into 
a series of subbands with increasing energies as the electron motion is being limited. 
Since the superconducting properties strongly depend on the density of states around
the Fermi surface, the superconducting gap of a metallic nanostructure drastically changes
each time when  the subband passes though the Fermi level. The size-dependent enhancement of the
energy gap induced by the quantum size effect has been theoretically
investigated by Shanenko et al. in Refs.~\cite{Shanenko2006, Shanenko2006_2}, 
for Al and Sn nanowires. Within these studies the experimentally observed width-dependent increase of $T_C$ for 
Al nanowires~\cite{Zgirski2005,Savolainen2004} has been reproduced.

The oscillations of the supercondcuting gap as a function of the thickness in ultrathin nanofilms were predicted
by Blatt and Thomson in 1963~\cite{Blatt1963}. Although it was expected that the quantum size effect in the superconducting 
nanofilms would not be an important factor (as the size reduction concerns only one dimension), recent experiments for 
Pb nanofilms~\cite{Guo2004,Ozer2006} grown on a Si(111) substrate
have shown, that the critical temperature and the critical magnetic field oscillate as a function of the
nanofilm thickness. Furthermore, it has been found that the superconductivity of the nanofilm
is not destroyed by the fluctuations even if the thickness of the nanofilm is only a single monolayer~\cite{Zhang2010}.
The oscillations of the superconducting gap as a function of the thickness for Pb nanofilm have been studied
theoretically in Ref.~\cite{Chen2013}. 
In these considerations the effective Fermi level has been introduced in order to reproduce the experimental results for 
the nanofilms~\cite{Chen2013} and nanowires~\cite{Shanenko2006, Shanenko2006_2}.
Such fitting method leads to a situation in which the effective Fermi level used in the calculations is almost an order 
of magnitude smaller than the one measured in the bulk. 
This means that the electron density in the nanofilm was reduced by a few orders of magnitude as compared 
to the electron density usually measured in metals.

In the present paper, we study the influence of the electron density on the thickness-depend superconducting gap
oscillations for the metallic nanofilms. Firstly, we have carried out calculations with the effective Fermi level and shown that 
the quantum size effect leads to the oscillations of the order parameter 
as a function of the nanofilm thickness. For sufficiently thin nanofilms, with thickness of 1-2 nm, the order parameter reaches 
value which is five times higher than the one corresponding to the bulk. In the next step, we have carried out the calculations 
for different electron densities (Fermi levels) and found that the amplitude of the energy gap oscillations decreases with 
increasing electron density. We have shown that the amplitude reduction is significant for the electron density corresponding to that measured in metals.
Finally, the thickness-dependent oscillations of the superconducting gap are also studied 
as a function of the electron-phonon coupling.

The present paper is organized as follows.
In Sec.~\ref{sec:model} we introduce the basic concepts of the calculation
scheme based on the BdG equations.
In Sec.~\ref{sec:results} we analyze the results, while the
conclusions and summary
are included in Sec.~\ref{sec:concl}.

\section{Theoretical method}\label{sec:model}
The microscopic theory based on the Bogoliubov-de Gennes (BdG) equations is
a natural way to describe the superconducting properties of nanofilms where
the quantum size effect entails the position-dependent order parameter $\Delta(z)$. 
The BdeG equations have the form
\begin{equation}
\left (
\begin{array}{cc}
-\frac{\hbar^2}{2m} \nabla ^2-\mu & \Delta(\mathbf{r}) \\
\Delta(\mathbf{r}) & \frac{\hbar^2}{2m} \nabla ^2+\mu
\end{array}
\right ) 
\left (
\begin{array}{c}
\mathcal{U}_i(\mathbf{r}) \\
\mathcal{V}_i(\mathbf{r})
\end{array}
\right )=
E_i
\left (
\begin{array}{c}
\mathcal{U}_i(\mathbf{r}) \\
\mathcal{V}_i(\mathbf{r})
\end{array}
\right ),
\label{BdG_start}
\end{equation}
where $\mathcal{U}_i(\mathbf{r})$ and $\mathcal{V}_i(\mathbf{r})$ are the electron-like and hole-like wave 
functions, $E_i$ is the quasi-particle energy, $m$ is the free electron mass,
$\mu$ is the chemical potential, 
and $\Delta(\mathbf{r})$ is the position-dependent order parameter, which in
the absence of the magnetic field, is a real quantity.\\
Assuming the periodic boundary conditions
in the $x-y$ plane, the quasi-particle wave functions can be expressed as
\begin{equation}
\left (
\begin{array}{c}
\mathcal{U}_{k_xk_y\nu}(\mathbf{r}) \\
\mathcal{V}_{k_xk_y\nu}(\mathbf{r})
\end{array}
\right ) = 
\frac{e^{ik_xx}}{\sqrt{L_x}} \frac{e^{ik_yy}}{\sqrt{L_y}}
\left (
\begin{array}{c}
u_{\nu}(z) \\
v_{\nu}(z)
\end{array}
\right ).
\label{WV}
\end{equation}
In the equation written above, the index $i$ has been replaced by $\{ k_x,k_y,\nu\}$, where $k_x$, $k_y$ are the free
electron wave vector components in the $x$ and $y$ direction, respectively while $\nu$ labels the subsequent quantum states in the
$z$ direction.\\
By substituting the wave function given by (\ref{WV}) into the BdG equations we
obtain 
\begin{eqnarray}
 \label{BdG1Da}
 &&\left ( -\frac{\hbar^2}{2m} \frac{d^2}{dz^2} -\mu +\frac{\hbar^2 k^2_{\parallel}}{2m} \right ) u_\nu(z) + \Delta(z) v_\nu(z) = E_\nu u_\nu(z), \\
 \label{BdG1Db} 
 &&\left ( \frac{\hbar^2}{2m} \frac{d^2}{dz^2} +\mu -\frac{\hbar^2 k^2_{\parallel}}{2m} \right ) v_\nu(z) + \Delta(z) u_\nu(z) = E_\nu v_\nu(z), \\ 
\end{eqnarray}
where $k^2_{\parallel}=k_x^2+k_y^2$. \\ 
If we assume that the system is infinite in the $x$ and $y$ direction ($L_x, L_y \rightarrow \infty$), 
the order parameter $\Delta(z)$ can be expressed in the following manner
\begin{equation}
\Delta(z)=\frac{g}{2 \pi} \int d k_{\parallel} \: k_{\parallel} \sum _{\nu} u_\nu(z)v_\nu^*(z) \left [ 1-2f(E_\nu) \right ]
\label{delta},
\end{equation}
where $g$ is the electron-phonon coupling and $f(E)$ is the Fermi-Dirac distribution. 
The summation in Eq.(\ref{delta}) is carried out only over these states for which the
single-electron energy $\xi _{k_xk_y\nu}$
satisfies the condition  $\left | \xi _{k_xk_y\nu} \right | < \hbar \omega _D$,
where $\omega _D$ is the Debye 
frequency and $\xi _{k_xk_y\nu} $ is given by 
\begin{eqnarray}
 \xi _{k_xk_y\nu} &=& \int dz \bigg [ u_\nu^*(z) \left ( -\frac{\hbar^2}{2m} \frac{d^2}{dz^2} -\mu +\frac{\hbar^2 k^2_{\parallel}}{2m} \right ) u_\nu(z) \nonumber \\
 &+& v_\nu^*(z) \left ( -\frac{\hbar^2}{2m} \frac{d^2}{dz^2} -\mu +\frac{\hbar^2 k^2_{\parallel}}{2m} \right ) v_\nu(z) \bigg ].
 \label{ekin}
\end{eqnarray}
The system of equations (\ref{BdG1Da})-(\ref{BdG1Db}) and equation (\ref{delta})
are solved in a self consistent manner by using the following procedure:
in the first step, we find the quasi-particle wave functions by numerically
solving the BdG equations (\ref{BdG1Da})-(\ref{BdG1Db}) with the order parameter $\Delta(z)$ (in the first iteration 
we use $\Delta(z)=\Delta_{bulk}$, where $\Delta_{bulk}$ is the energy gap in the bulk).
In the next step, after inserting the quasi-particle wave functions into Eq.~(\ref{delta}),
we calculate the new order parameter profile $\Delta (z)$.
Using this profile, we again solve the BdG equations (\ref{BdG1Da})-(\ref{BdG1Db}).
This procedure is repeated until the convergence is reached. \\
Since the chemical potential for the nanostructures deviates from the bulk
value, for each nanofilm thickness 
we determine the chemical potential by using the formula
\begin{equation}
 n_e=\frac{1}{\pi d} \int dk_{\parallel} \: k_{\parallel} \sum _{\nu} \int dz \bigg [ |u_\nu(z)|^2f(E_\nu) +|v_\nu(z)|^2 (1-f(E_\nu))\bigg ],
\end{equation}
where $d$ is the thickness of the film in the $z$ direction, in which we use
the hard-wall potential profile leading to the 
boundary conditions $u_\nu(0)=u_\nu(d)=0$ and $v_\nu(0)=v_\nu(d)=0$.

\section{Results and Discussion}\label{sec:results}

In this section we present the results of calculations for Al nanofilms. The calculations have been carried out for 
the following values of the parameters: $gN_{bulk}(0)=0.18$ where $N_{bulk}(0)=mk_F/(2 \pi^2 \hbar ^2)$ 
is the bulk density of the single-electron states at the Fermi level, $\hbar \omega _D=32.31$~meV 
and the bulk energy gap $\Delta_{bulk}=0.25$~meV. 
According to Ref.~\cite{Shanenko2007}, the Fermi level in the bulk $\mu _{bulk}$ is treated as a fitting parameter and its value is determined based on the 
experimental results from the photoelectron spectroscopy i.e. it is taken on $\mu _{bulk}=0.9$~eV, which corresponds 
to the electron density $n_e \approx 4 \times 10^{21}$~cm$^{-3}$~\cite{Shanenko2006}.
The electron density associated with the effective Fermi level, is an order of magnitude lower 
as compared to the one measured in Al bulk  $n_e=1.8 \times 10^{23}$~cm$^{-3}$~\cite{Ashcroft}.
As it has been stated by the authors of Ref.~\cite{Shanenko2007}, this discrepancy results from the parabolic band approximation 
used in the model. 
Nevertheless, in our opinion the extended study of the influence of the electron density on the superconducting properties of metallic nanofims 
is needed and according to our knowledge, has not been reported until now.
For this purpose, we firstly present the results of calculations carried out for the effective Fermi level and then present how the 
considered phenomena are changed with increasing electron density up to the value measured in the bulk.

\begin{figure}[ht]
\begin{center}
\includegraphics[scale=0.3]{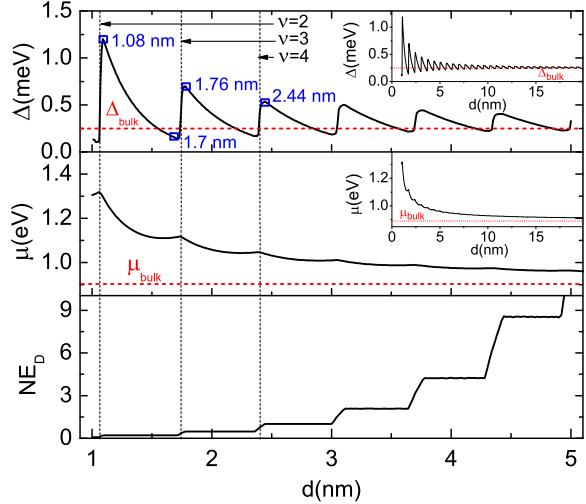}
\caption{(a) Superconducting energy gap $\Delta$, (b) chemical potential $\mu$ and (c) electron density of states in the energy 
window $E_D=\left [ \mu - \hbar \omega _D,  \mu + \hbar \omega _D \right ]$  as a function of the 
nanofilm thickness $d$. Internal panels present the thickness-dependencies $\Delta(d)$ and $\mu(d)$ in wider 
range of $d$ varying from $1$~nm up to $20$~nm and clearly show that both these parameters converge to their bulk 
values as the nanofilm thickness increases.}
\label{fig1}
\end{center}
\end{figure}

\begin{figure*}[ht]
\begin{center}
\includegraphics[scale=1.0]{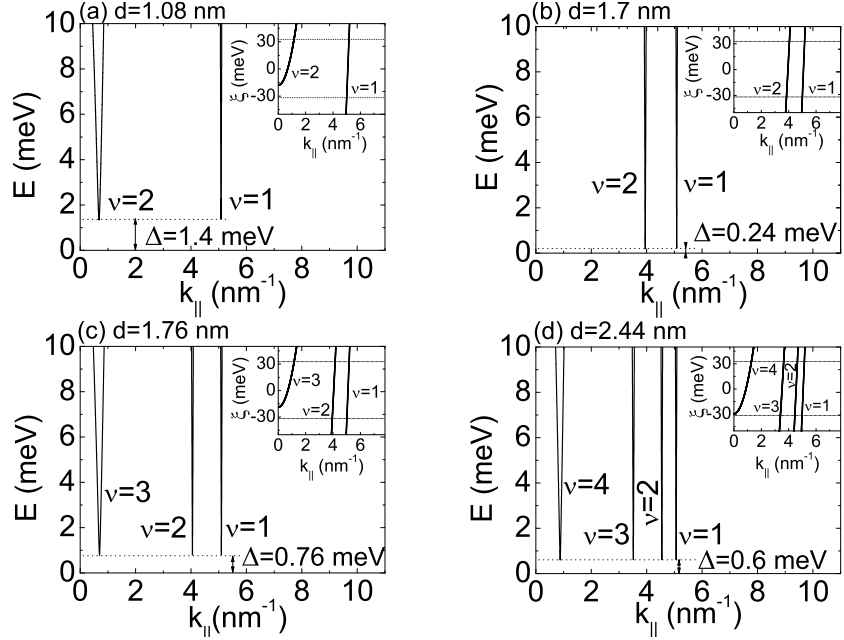}
\caption{Quasi-particle energy $E$ as a function of the wave vector $k_{||}$ for nanofilm thicknesses (a) $d=1.08$~nm, 
(b) $d=1.7$~nm, (c) $d=1.76$~nm and (d) $d=2.44$~nm [see Fig.~\ref{fig1}(a)]. The internal panels display the kinetic energy 
$\xi$ vs $k_{||}$. The energy window $\left [ \mu - \hbar \omega _D,  \mu + \hbar \omega _D \right ]$  in the internal panels are marked by
dashed horizontal lines.}
\label{fig2}
\end{center}
\end{figure*}

\begin{figure}[ht]
\begin{center}
\includegraphics[scale=0.25]{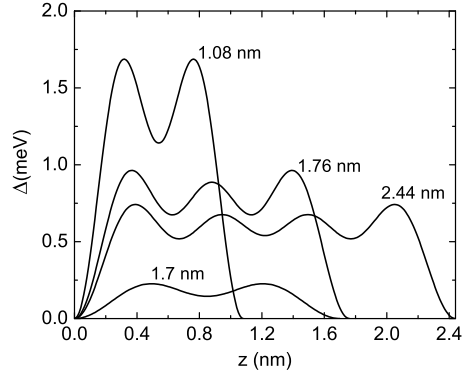}
\caption{Position-dependent energy gap $\Delta(z)$ for nanofilm thicknesses marked by
squares in Fig.~\ref{fig1}(a).}
\label{fig3}
\end{center}
\end{figure}

In Fig.~\ref{fig1}(a) we present the superconducting  parameter $\Delta$ as a function of the nanofilm thickness $d$ 
calculated for the effective Fermi level $\mu _{bulk}=0.9$~eV.
The dependence $\Delta(d)$ shows that for some particular value of $d$ the energy gap abruptly increases 
reaching the value about 5 times higher as compared to $\Delta_{bulk}$ [Fig.\ref{fig1}(a)]. 
The oscillations of the energy gap as a function of the nanofilm thickness 
correspond to the confinement of the electrons in the nanofilms and can 
be understood as follows.
In the superconducting state the Cooper-pairs are formed by electrons with energies from the range close to the Fermi level. 
This energy range is determined by the electron-phonon coupling and is limited by the Debye energy
$\hbar \omega _d$, where $\omega _d$ is the Debye frequency.
It means that the superconducting gap strongly depends on the number of 
states in the energy window $\left [ \mu - \hbar \omega _D,  \mu + \hbar \omega _D \right ]$ around the Fermi level.
In the ultra thin nanofilm, the electron motion in the direction perpendicular to the surface is limited to the nanometer scale 
what leads to the quantization of the electron energy. In the free electron model, the Fermi sphere transforms into the series of the 
parabolic subbands which position on the energy scale decreases with increasing nanofilm thickness.
If we increase the thickness $d$, the subsequent subbands pass through the energy window 
$\left [ \mu - \hbar \omega _D,  \mu + \hbar \omega _D \right ]$ which leads to the step-like enhancement of the density of states 
participating in the condensation of the Cooper-pairs [see Fig.~\ref{fig1}(c)].
The described mechanism is responsible for the increase of the energy gap $\Delta$ depicted in Fig.~\ref{fig1}(a). 
As one can see the peaks in the $\Delta(d)$ dependence are accompanied by a small chemical potential increasement 
[see Fig.~\ref{fig1}(b)]. In Fig.~\ref{fig1}(a) the highest enhancement of the energy gap is observed for 
the first three maximums, which correspond to the condensation 
of Cooper-pairs from the second, third, and forth subband, respectively. 
In Fig.~\ref{fig2} we present the quasi-particle 
energy $E$ and the kinetic energy $\xi$ (internal panels) as a function of the wave vector $k_{||}$ for several nanofilm thicknesses: 
(a) $d=1.08$~nm  which corresponds to the first maximum of $\Delta(d)$, (b) $d=1.7$~nm which corresponds to the drop of the energy gap below its bulk value, 
(c) $d=1.76$~nm and (d) $d=2.44$~nm which correspond to the second and third maximum of $\Delta(d)$.
The thicknesses for which  $E$ vs $k$ and $\xi$ vs $k$  dispersions have been calculated, are marked by squares in Fig.~\ref{fig1}(a).
Wee can see that for the nanofilm thickness $d=1.08$~nm [Fig.\ref{fig2}(a)] the enhancement of the energy gap corresponds to the Cooper pairing 
of electrons from the quantum subband $\nu=2$ which kinetic energy minimum is located in the energy window $\left [ \mu - \hbar \omega _D,  \mu + \hbar \omega _D \right ]$.
By analogy, the analysis of Figs.~\ref{fig2}(c) and (d) allows us to conclude that the second and third maximum correspond to the condensation of electrons from 
the subband $\nu=3$ and $\nu=4$, respectively. In contrary, in Fig.~\ref{fig2}(b) we can observe that the drop of the energy gap below 
its bulk value results from the fact that the minimum of the subband $\nu=2$ leave the energy window $\left [ \mu - \hbar \omega _D,  \mu + \hbar \omega _D \right ]$.
The participation of the subsequent subbands in the creation of the superconducting state leads also to the inhomogeneity of the energy gap in the $z$ direction 
which is presented in Fig.~\ref{fig3}. In this figure the number of maximums corresponds to the state number $\nu$ which is 
responsible for the enhancement of the energy gap.
It should be noted, that the amplitude of $\Delta(d)$ oscillations in Fig~\ref{fig1}(a) decreases with increasing nanofilm thickness, for which the higher 
excited states participates in the Cooper pair condensation.
The internal panels of Fig.~\ref{fig1} presents the thickness-dependencies $\Delta(d)$ and $\mu(d)$ in wider 
range of $d$ varying from $1$~nm up to $20$~nm and clearly show that both $\Delta$ and $\mu$ converge to their bulk 
values as the nanofilm thickness increases.

\begin{figure}[ht]
\begin{center}
\includegraphics[scale=0.3]{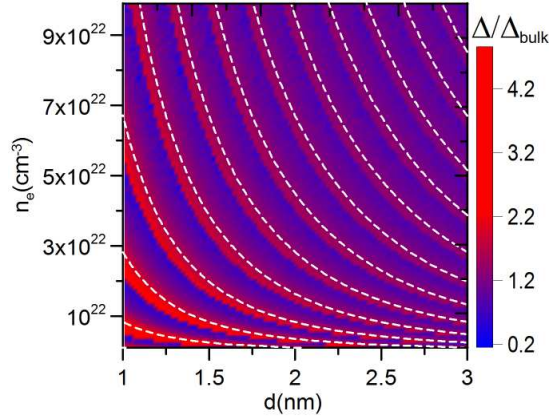}
\caption{Superconducting energy gap $\Delta$ as a function of electron density $n_e$ and nanofilm thickness $d$. White dashed 
lines mark the nanofilm thicknesses estimated from single-electron energy level, for which the subsequent
quantum states pass through the Fermi level.}
\label{fig4}
\end{center}
\end{figure}

As it can be seen, within the considered model, with the effective Fermi level, one can predict the appearance of the energy gap oscillations 
as a function of nanofilm thickness and almost five-fold enhancement of the energy gap for proper value of $d$. However, these predictions are significantly weakened
if we increase the electron density up to the value measured for Al ($n_e=1.8 \times 10^{23}$~cm$^{-3}$). In Fig.~\ref{fig4} 
we present the energy gap as a function of the nanofilm thickness and the electron density varying
from $10^{21}$~cm$^{-3}$ up to $10^{23}$~cm$^{-3}$. We restrict our analysis to the thickness range $1-3$~nm for which 
the highest maximums of $\Delta(d)$ are observed in Fig.~\ref{fig1}(a).
Fig.~\ref{fig4} shows that the increase of the electron density results in the decrease of the period of $\Delta(d)$ oscillations.
This behavior can be explained in terms of the quantization of the quasi-particle energy levels in the direction perpendicular to the plane.
The energy level in the presence of the Cooper pairing can be well estimated by the single-electron energy level by  
using the formula $E \approx \hbar ^2 \pi ^2 \nu^2 /(2md^2)$, where the hard-wall potential is assumed in the $z$ direction.
It means that the quantum state $\nu$ passes through the Fermi level for the nanofilm thickness 
$d \approx \hbar \pi \nu / \sqrt[3]{3 \pi^2 n_e}$.
The distance between two neighboring peaks can be estimated by $\Delta d =  \pi / \sqrt[3]{3 \pi^2 n_e}$.
In Fig.~\ref{fig4}, the estimated nanofilm thicknesses for which the subsequent 
quantum states pass through the Fermi level are marked by white dashed lines. We see that the
 single-electron energy level approximation can well reproduced the position of the energy gap peaks in the
 dependence $\Delta (d,n_e)$.

\begin{figure}[ht]
\begin{center}
\includegraphics[scale=0.25]{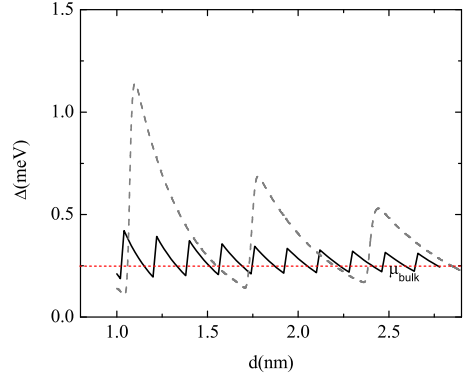}
\caption{Superconducting energy gap $\Delta$ as a function of nanofilm thickness $d$ calculated for electron 
density $n_e=1.8 \times 10^{23}$~cm$^{-3}$ corresponding to the bulk value for Al. For comparison $\Delta(d)$ calculated
for the effective Fermi level is displayed by the dashed gray line.}
\label{fig5}
\end{center}
\end{figure}

The most important feature which can be found in Fig.~\ref{fig4} is the decrease of the amplitude of
$\Delta(d)$ oscillations with increasing electron density. 
Fig.~\ref{fig4} depicts that if we increase the electron concentration, the maximum enhancement 
of the energy gap decreases from almost five-fold (as compared to $\Delta_{bulk}$) for $n_e=10^{21}$~cm$^{-3}$, to 
less than twice for the electron density $n_e= 10^{23}$~cm$^{-3}$.
This fact allow us to conclude that the significant enhancement 
of the energy gap as a function of the nanofilm thickness can be observed only for the superconductor with the
low concentration of carriers, e.g., SrTi0$_3$ which exhibits superconductivity in the carrier concentration regime
$10^{18}-10^{19}$~cm$^{-3}$~\cite{Schooley}. The fact that $\Delta(d)$ oscillations are less pronounced in the high-carrier
concentration materials is crucial with respect to the experimental observation of the considered phenomena.
In Fig.~\ref{fig5} we present $\Delta(d)$ calculated for the electron density $n_e=1.8 \times 10^{23}$~cm$^{-3}$ corresponding
to the bulk value for Al. This figure shows that for such high electron density the oscillations of the energy gap as a function of
the nanofilm thickness are significantly suppresed. 

\begin{figure}[ht]
\begin{center}
\includegraphics[scale=0.8]{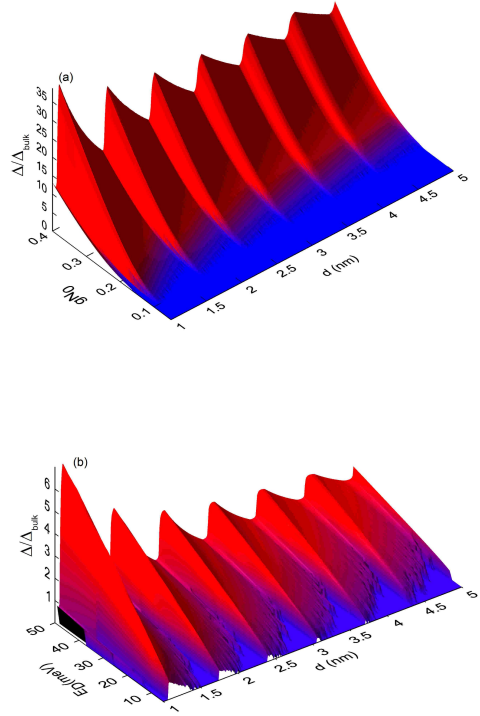}
\caption{
Thickness-dependent energy gap oscillations as a function of (a) the electron-phonon coupling $gN(0)$
and (b) the Debye energy $\hbar \omega _D$. Calculation carried out for the effective Fermi level $\mu _{bulk}=0.9$~eV.
}
\label{fig6}
\end{center}
\end{figure}

The microscopic model based on BdG equations  allows to determine the parameters 
describing the superconducting state in metallic nanofilms when its thickness is reduced to few nanometers.
However, in the nanoscale regime the superconductivity of the nanofilms is changed not only by the quantization of the 
quasi-particle energy in the confining direction, but also due to the fact that the phonon modes in the nanostructure
(which mediate the Cooper pairing) strongly deviate from that observed in the bulk. Such deviation was experimentally reported 
for Ag nanofilm on the Fe substrate~\cite{Luh2002}. 
Therefore, the assumption that the electron-phonon coupling is constant and equal to the bulk value 
is the weak point of the presented considerations. 
Since the calculation of the electron-phonon coupling in the ultra thin nanofilms requires the \textit{ab initio} method we
treat $gN(0)$ as well as the Debye energy $\hbar \omega _D $ as parameters and calculate how 
the energy gap in the nanofilms changes as a function these two quantities [Fig.~\ref{fig6}(a) and Fig.~\ref{fig6}(b)].
Fig.~\ref{fig6} shows that for each value of the nanofilm thickness the energy gap  
is an increasing function of the parameters $gN(0)$ and $\hbar \omega _D$.
Therefore, one can expect that the experimentally observed decrease of the electron-phonon coupling 
as a function of the nanofilm thickness results in the decrease of the amplitude of $\Delta(d)$ oscillations in the 
metallic nanofilm, which is the second factor in addition to the electron concentration which 
weakens the energy gap oscillations effect. It is also worth to note, that although the values of $\Delta(d)$ increase
as a function of the parameters $gN(0)$ and $\hbar \omega _D$, the ratio between the maximum and the minimum 
value of the energy gap for each peaks remains almost unaffected by the changes in $gN(0)$ and $\hbar\omega$, 
e.g. it is about five for the first peak.

\section{Conclusions}\label{sec:concl}

The influence of the electron density on the oscillations of 
the superconducting energy gap as a function of the nanofilm thickness 
has been studied based on the self-consistent numerical solution of the Bogoliubov-de Gennes equations.
We have shown, that in the ultra thin nanofilms, the strong enhancement of the superconducting energy gap for particular nanofilm thicknesses corresponds to 
the quasi-particle energy quantization induced by the confinement of electrons in the direction perpendicular to the film.
In such situation, the Fermi sphere transforms into a series of parabolic subbands with energies that are decreasing with increasing nanofilms thickness.
Each time when the subband passes through the energy window 
$\left [ \mu - \hbar \omega _D,  \mu + \hbar \omega _D \right ]$, the enhancement of the energy gap occurs in 
the dependence $\Delta(d)$.
In the present paper we have studied the influence of the electron density on the thickness-dependent energy gap oscillations in Al nanofilms.
We have found that the amplitude of the $\Delta(d)$ oscillations decreases with increasing electron density. 
The calculations carried out for the electron concentration corresponding to the one measured in the Al bulk
have shown that $\Delta(d)$ oscillations are significantly reduced and almost completely disappear. 
This fact is relevant with respect to the experimental observation of the 
energy gap oscillations in the superconducting nanofilms. It allows us to restrict the experimentaly investigation 
to the low-carrier concentration superconductor e.g., SrTi0$_3$.
Furthermore, we have shown that the period of  $\Delta(d)$ oscillations is a decreasing function of the electron concentration.
This behavior and the positions of the maximums in the dependence $\Delta(s,n_e)$ have been well reproduced by the use of the single-electron 
energy level approximation.
Since the reduction of the dimensionality changes considerably the phonons dispersion, we have also studied the influence of the 
electron-phonon coupling and the Debye energy on the superconducting energy gap oscillations. These studies show that the decrease of 
the electron-phonon coupling observed experimentally in ultra thin nanofilm entails the decrease of the amplitude of $\Delta(d)$ oscillations.

Summing up, our results show that the  influence of the changes in the electron density and the electron-phonon coupling constant on the magnitude 
of the considered phenomena is strong and should be taken into account in the theoretical investigations concerning superconductivity in  metallic nanofilms.  
In particular, the increase of $n_e$ and the decrease of g parameters leads to a relevant reduction of the oscillations amplitude, which could explain 
why in various experiments the considered phenomena hasn't been clearly visible. At the end, it is worth mentioning, that the matter of the nanofilm thickness 
dependance of the electron-phonon coupling constant is still not completely settled and with this respect, a proper $ab$ $initio$ calculations could make a 
substantial contribution in the complete theoretical description of the superconducting state in the nanofilm samples.

\section*{Acknowledgements}\label{sec:aments}
Discussions with J{\'o}zef Spa{\l}ek are gratefully acknowledged.
This work  was financed from the budget for Polish Science in the years 2013-2014. Project number: IP2012 048572.
M. Z. acknowledges the financial support from the Foundation for Polish Science (FNP) within project TEAM.

\section*{References}
\bibliographystyle{unsrt.bst}
\bibliography{refs}
\end{document}